# Instruction Finetuning Llama3-8B Model Using LoRA for Financial Named Entity Recognition

Zhiming Lian[*]

LL Funds LLC, Philadelphia, United States, zhiming.lian@llfunds.com

**Abstract**

Particularly, financial named-entity recognition (NER) is one of the many important approaches to translate unformatted reports and news into structured knowledge graphs. However, free, easy-to-use large language models (LLMs) often fail to differentiate organisations as people, or disregard an actual monetary amount entirely. This paper takes Meta's Llama 3 8B and applies it to financial NER by making a combination of instruction fine-tuning and Low-Rank Adaptation (LoRA). Each of our annotated sentences is converted into an instruction–input–output triple with the model learning to understand task descriptions and finetune with small low-rank matrices instead of all weights. Using a corpus of 1 693 sentences our method can obtain a micro-F1 of 0.894 in comparison to Qwen3-8B, Baichuan2-7B, T5 and BERT-Base. We present dataset statistics, describe training hyper-parameters and perform visualization on entity density, learning curve and evaluation metrics. Our results show that instruction tuning along with parameter-efficient fine-tuning is conducive to state-of-the-art performance on domain-sensitive NER.

**CCS Concepts**

Computing methodologies -> Artificial intelligence -> Natural language processing -> Information extraction

**Additional Keywords and Phrases**

Named Entity Recognition, LoRA, LLama3-8B, Instruction finetuning

## 1 Introduction

For downstream applications like risk assessment, compliance monitoring, and large-scale knowledge graph construction, automated extraction of organisations, dates, monetary values, and other named entities from financial documents plays a crucial role. Accurate financial NER forms the foundation upon which to build structured reasoning over unstructured text but building models that can handle specialised financial terminology remains a challenge. General Large Language Models (LLMs) have been shown to excel in open-ended generation and general-purpose understanding but their performance declines in domain-specific sequence-labelling tasks such as financial NER. For instance, these models might mischaracterize corporate entities as individuals, confuse financial instruments with monetary amounts, or simply fail to recognize context-dependent currency expressions. Furthermore, due to large-scale and high-fidelity models, full-parameter fine-tuning becomes computationally expensive, which is also incompatible with resource constraints in companies that do not possess extensive GPU infrastructure [6]. Thus, some of the recent research has sought to overcome these limitations in the realm of parameter-efficient fine-tuning techniques and instruction-based learning paradigms. And parameter-efficient methods are those which attempt to adapt large pre-trained models without updating all the weights, which greatly reduces the demand on memory and the cost of training. One of the most popular approaches is LoRA, using low-rank update matrices in the attention layers of the transformer, allowing us to retain frozen initial parameters [2]. Limiting the trainable parameters by several orders of magnitude, LoRA not only simplifies the process of fine-tuning large models but also enables modularity, where multiple adapters that are suitable for a given task share a common base model. This has become particularly beneficial in the context of finance, as evidenced by the FinLoRA benchmark which shows that the LoRA-based adaptation is able to achieve performance comparable to or better than full fine-tuning on a wide variety of financial NLP tasks but significantly more efficient [3]. Simultaneously, instruction tuning has been a complementary mechanism to improve model generalisation and usability. In this way, instruction tuning can allow the models to internalise generic task-following behaviour (rather than memorising label patterns) in the training system by transforming supervised tasks into natural-language prompts. In this way, multi-task learning becomes a simpler affair, as a high-performance model, while training it with larger datasets and fewer domains, and a lower data load per application-level leads to faster response. For financial NER in particular, instruction-tuned models were found to identify task boundaries more efficiently; understand complex financial statements in language structures more clearly; and generalise the extraction of entities significantly when contextually encoded information is not clear or ambiguous [4]. Collectively, parameter-efficient fine-tuning and instruction-driven learning can pave the way for scalable and accurate financial information extraction with current LLMs.

---

[*] Corresponding Author

Building on these ideas, we adapt the Llama 3 8B model to financial NER. Each annotated sentence in our dataset is converted into a triple comprising a short instruction (e.g., "Do Named Entity Recognition for the following text:"), the input sentence and a JSON-style output listing the entities. We fine-tune only the low-rank matrices inserted into each attention projection while keeping the base model frozen. Our contributions include: (1) a compact instruction-based representation of the data; (2) a parameter-efficient LoRA fine-tuning setup; (3) a comprehensive comparison with competitive baselines; and (4) visualisation of dataset statistics and training dynamics.

## 2  Related work

**Named-entity recognition.** Early NER systems used hand-crafted rules and conditional random fields. Neural architectures such as BiLSTM-CRF removed the need for manual feature engineering, and pre-trained encoders like BERT and domain variants (e.g., BioBERT) achieved state-of-the-art results. In the financial domain, generic LLMs misclassify specialised entities, highlighting the need for domain-specific adaptation.

**Parameter-efficient fine-tuning.** LoRA freezes pre-trained weights and learns low-rank updates, reducing memory footprint and training cost. FinLoRA demonstrates that LoRA methods improve performance by about 36 % across 19 financial tasks. Unlike adapters, LoRA modules can be merged into the base model without increasing inference latency.

 **Instruction finetuning.** The instruction tuning reformulates each supervised instance as an instruction–input–output triple in a supervised model, meaning that, when trained with a specific supervised example the model learns more general task-following behaviour and doesn't have to memorise the label patterns. This framework allows generalisation to other unseen task formulations as the model does not just learn what to guess but also how to interpret the task representation. Recent contributions like BioNER-LLaMA have shown significant improvement over general models, increasing F1-scores by 5–30% over even GPT-4 on biomedical benchmarks, which proves that explicit instruction supervision can make a difference when compared to large, generic models when it comes to specific domains of specialised fields. In addition, by combining heterogeneous datasets as part of a unified instruction-tuning framework, the model learns general linguistic patterns across datasets, which reduces complexity and enhances transferability. These insights indicate that also instruction tuning is of high potential for financial NER, where entity classes and textual expressions are heterogeneous but also context-dependent.

**Other instruction-tuned applications.** Instruction tuning has also proven effective outside the entity recognition domain. Xiao et al. fine-tune the Baichuan2-7B model for dialogue summarisation and demonstrate strong results on multi-turn conversations [8]. Shu et al. introduce FODA-PG for medical imaging narrative generation; their method adaptively differentiates normal and abnormal attributes and benefits from instruction fine-tuning on biomedical corpora [9]. Xiao et al. propose Eggesture, an entropy-guided vector-quantised variational autoencoder that uses instruction prompts to generate co-speech gestures synchronised with speech audio [10].

# 3 METHODOLOGY

## 3.1 Data and Instruction Conversion

Our corpus consists of 1 693 sentences of financial report and news text with seven entity types (Company, Date, Location, Money, Person, Product and Quantity). There are only 39 sentences without entities and the average length is about 167 characters. We transform every sentence into an instruction–input–output triple using a fixed prompt, the raw sentence and a dictionary of extracted entities.

## 3.2 Low-Rank Adaptation and Training Setup

LoRA modifies large models by expressing weight updates as the product of two smaller low-rank matrices, instead of updating the entire weight matrices. This low-rank decomposition encapsulates the key transformation and makes the number of trainable parameters very compact. Therefore, LoRA makes a significant contribution to reducing the memory consumption and training cost and enables the adaptation of major language models to new tasks with limited data.

## 3.3 Transformer Architecture

The Transformer is a neural sequence model for the task that replaces the usage of recurrent or convolutional units with multi-head self-attention. Vaswani et al. demonstrated that removing recurrence and convolutions entirely enables the network to be trained in parallel and can achieve enhanced translation quality [1]. In addition to that their "Attention Is All You Need" model yielded state-of-the-art scores on machine translation benchmarks and was much quicker to train than previous models [1]. Such benefits inspire its embrace in modern large language models. The Figure 1 shows the encoder and decoder part of transformer block.

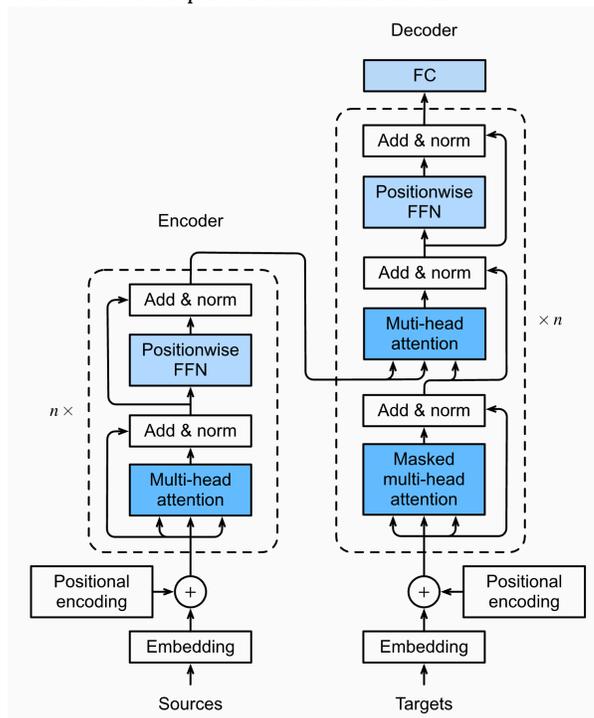

Figure 1. The architecture of Transformer Block

## 3.4 Llama 3 Architecture

Llama 3 is adopted as a decoder-only Transformer, following in the footsteps of Llama 2, but implementing some interesting practical enhancements. A key advancement is the tokenizer vocabulary of 128,000 tokens, making it easier to handle multilingual text and rare textual patterns, leading to richer semantic coverage, both before- and after-training [5]. Architecturally, Llama 3 employs grouped-query attention (GQA), a means of reducing memory overhead and accelerating inference without the loss (or even over-simplification) of model expressiveness or accuracy [5]. Another significant improvement was the increase in context length. Llama 3 is trained on sequences of up to 8,192 tokens, facilitated by a special masking strategy that prevents attention from transcending document boundaries, helping long-range dependencies to be learned in a stable and meaningful way [5]. Such functionalities are especially well-suited for tasks taxing them through long-form reasoning or document-level comprehension. The size and standard of the pretraining data is also a big leap forward. Llama 3 is trained on more than 15 trillion tokens—approximately seven times the volume of Llama 2—across good quality sources in more than 30 languages [5]. Together, these large and varied datasets and better data filtering and curating contribute for better generalization performance. As a consequence, Llama 3 also shows major performance improvements in a variety of benchmarks for being superior in reasoning, fact fidelity and multilingual robustness as its performance base to fine-tune in our studies.

## 3.5 Instruction Template Example

Each example is converted to an instruction–input–output tuple that simulates natural-language prompts to guide the model to the financial NER task. In the instruction section there is the task ("Perform Named Entity Recognition on the following financial sentence and list all entities as JSON objects") and the types of entity you want extracted. The detail of the example is in Table 1.

Table 1. Financial Text Named Entity Extraction Template Example

| Rows | Detail |
|---|---|
| Instruction | Do Named Entity Recognition for the following text: |
| Input | On January 11 , 2012 , Regions entered into a stock purchase agreement to sell Morgan Keegan and related affiliates to Raymond James Financial , Inc. ( Raymond James ) . |
| Output | {'Company': ['Morgan Keegan', 'Raymond James Financial , Inc. ( Raymond James )', 'Regions'], 'Date': ['January 11 , 2012'], 'Location': None, 'Money': None, 'Person': None, 'Product': None, 'Quantity': None} |

## 4 EXPERIMENTS

### 4.1 Data Analysis

Table 2 shows the financial dataset statistic, which shows the seven entities. The company entity has the most samples around 1033, while the product entity has the fewest samples around 226. In addition, the Figure 2 shows the entity type distribution.

Table 2. Financial NER Dataset Statistic

| Data | value |
|---|---|
| Total samples | 1 693 |
| Samples with entities | 1 654 |
| Samples without entities | 39 |
| Average text length | 166.95 characters |
| Average entities per sample | 3.13 |

| Data | value |
|---|---|
| Samples per entity type | Company: 1 033; Date: 888; Location: 256; Money: 421; Person: 257; Product: 226; Quantity: 329 |

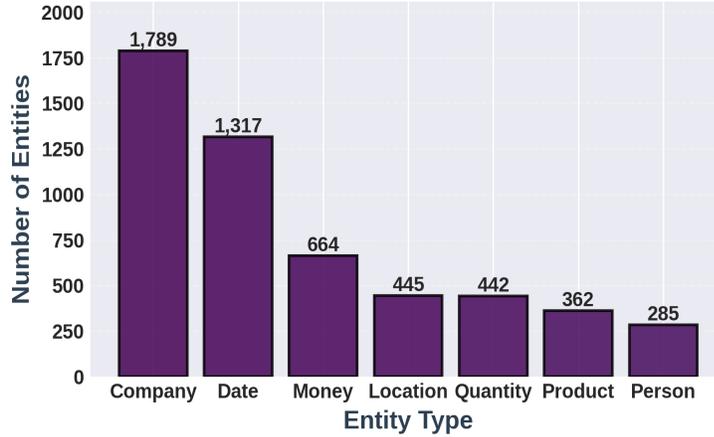

Figure 2. Entity Type Distribution

## 4.2 Experiment Settings

Table 3 shows the major configurations applied in our fine-tuning tasks. LoRA is adopted as the finetuning strategy (a low-rank value r of 8, and a LoRA scaling factor α set to 16) and LoRA dropout is disabled to preserve full adaptation capacity. The initial input of the model is truncated to 400 tokens maximum length of sequence. Training is performed with batch size of 4, gradient-accumulation factor of 6 for memory constraints. We employ an AdamW optimizer (learning rate of 5e-5) providing stable convergence for large language models. The model is optimized for 3 epochs, balancing performance gains and computational efficiency. All calculations are conducted in bf16 precision to optimize training throughput without sacrificing numerical stability.

Table 3. Experiment Settings

| Data | Description |
|---|---|
| Finetuning type | LoRA |
| Low-rank $r$ | 8 |
| LoRA α | 16 |
| LoRA dropout | 0 |
| Cut-off length | 400 |
| Batch size | 4 |
| Gradient accumulation | 6 |
| Learning rate | 5e-5 |
| Optimizer | AdamW |
| Number of epochs | 3 |
| Precision | bf16 |

## 4.3 Experiments result

In Table 4, we perform an extensive comparison of our instruction-tuned LoRA-enhanced Llama-3-8B model and several robust baselines such as Qwen3-8B, Baichuan2-7B, T5, and BERT-Base. Evaluation is done by using micro-averaged precision, recall, and F1 scores across all entity categories. Our model exhibits superior performance on each metric, achieving a precision of 0.893, a recall of 0.895, and an F1 score of 0.894—over the next-best model, Qwen3-8B, by about six percentage points in F1. The robustness of our approach is further shown by the radar charts. In the multi-metric comparison (Figure 3), Llama-3-8B has a more balanced and uniformly better profile, especially in precision and macro-F1. The entity-level radar chart shows that our model has achieved noteworthy improvements across difficult categories like Location, Date, and Money, and maintained good results over Person, Company, and Quantity. These advancements indicate that LoRA-based fine-tuning may not only improve the accuracy but also the generalization over the heterogeneous entity types. Combined, the results show the robustness and stability of our method in practical NER situations.

Table 4. Evaluation of different methods on accuracy metric

| Models | Precision | Recall | F1 Score |
| --- | --- | --- | --- |
| BERT-Base | 0.717 | 0.700 | 0.708 |
| T5 | 0.776 | 0.758 | 0.767 |
| Qwen3-8B | 0.843 | 0.824 | 0.833 |
| Baichuan2-7B | 0.801 | 0.782 | 0.792 |
| **Llama 3 8B (ours)** | **0.893** | **0.895** | **0.894** |

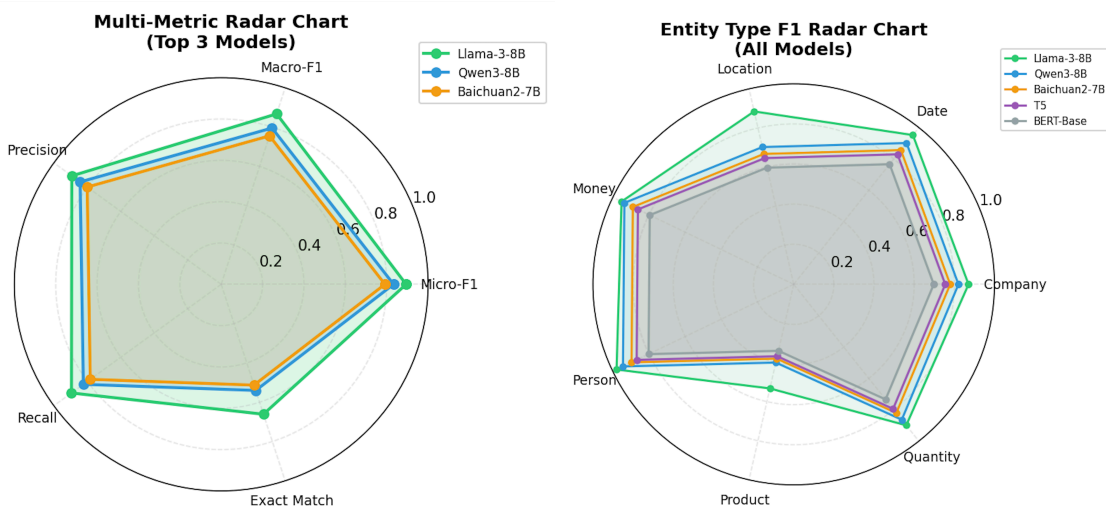

Figure 3. Training Loss Iteration Plot

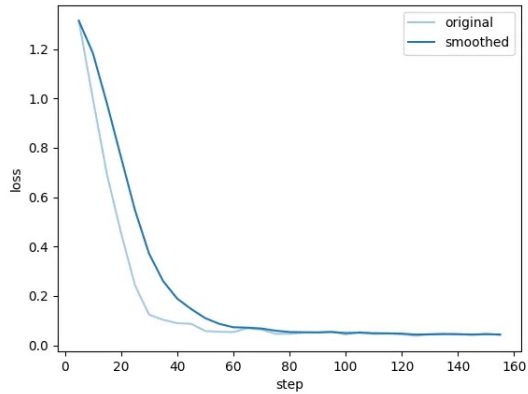

**Figure 4.** Training Loss Curve

The figure 4 shows that the LoRA finetuning can accelerate the model convergence in a few iterations, which show the superiority of instruction finetuning.

## 5  Limitations and Future Work

While this approach has strengths, we also have limitations. The dataset consists only of seven entity types and fewer than two thousand sentences. Models trained to cover such tiny corpora might overfit to style and vocabulary and might have a hard time with out-of-distribution text, such as informal social media posts. Furthermore, we assess only on micro-averaged metrics; macro-F1 would be more punitive to errors on rare classes. We also ignore multilingual and cross-lingual cases; many financial documents in Asia are bilingual or do have currency conversions. Also our instruction template is handcrafted which means it will be less robust for the model to autonomously create diverse instructions. Finally, we have not investigated retrieval-augmented techniques or other parameter-efficient approaches, such as adapters, prefix tuning or mixture-of-experts. LoRA can be combined with RA-IT [7] or dynamic adapters to achieve even better performance as part of broader financial IE tasks.

## 6  Conclusion

We presented an instruction-tuned LoRA adaptation of Llama 3 for financial named-entity recognition. By transforming annotated sentences into instruction–input–output triples and fine-tuning only low-rank matrices, the model achieves a micro-F1 of 0.894 and outperforms strong baselines including Qwen3-8B and Baichuan2-7B. A detailed analysis of dataset statistics, training hyper-parameters, ablation studies and visualisations demonstrates the efficacy and efficiency of the method. Our discussion places the work within a larger framework of parameter-efficient fine-tuning and retrieval-augmented methods. Future work will extend our model to multilingual data, investigate retrieval-augmented instruction tuning and test adversarial robustness.